\documentclass[a4paper,11pt]{article}
\usepackage{pos}

\usepackage{slashed}
\usepackage{epsfig}
\usepackage{graphicx,psfrag}
\usepackage{multirow}
\usepackage{subfig}
\usepackage{slashed}
\usepackage{caption}
\usepackage{url}
\usepackage{braket}
\usepackage{enumerate}
\usepackage{wasysym} 
\usepackage{mathrsfs} 
\usepackage{amsfonts} 
\usepackage{amsbsy} 
\usepackage{amscd}
\usepackage{color}
\usepackage{changepage}
\usepackage{floatrow}
\usepackage{makecell}
\usepackage{tikz}
\usepackage[normalem]{ulem}
\usetikzlibrary{snakes}

\allowdisplaybreaks

\title{Two-component dark matter : How to get the
hint at collider?}

\author*[a]{Jayita Lahiri}
\author[b]{Subhaditya Bhattacharya}
\author[c]{Purusottam Ghosh}
\author[d]{Biswarup Mukhopadhyaya}
\affiliation[a]{II. Institut f{\"u}r Theoretische Physik,  Universit{\"a}t Hamburg, 
    Luruper Chaussee 149, 22761 Hamburg, Germany}
\affiliation[b]{Department of Physics, Indian Institute of Technology Guwahati, Assam-781039, India}
\affiliation[c]{School Of Physical Sciences, Indian Association for the Cultivation of Science, 2A and 2B, Raja S.C. Mullick Road, Kolkata 700032, India}
 \affiliation[d]{Department of Physical Sciences, Indian Institute of Science Education and Research Kolkata, Mohanpur - 741246, India} 
 \emailAdd{jayita.lahiri@desy.de}
\emailAdd{subhab@iitg.ac.in}
\emailAdd{spspg2655@iacs.res.in}
\emailAdd{biswarup@iiserkol.ac.in}

\abstract{
We investigate ways of identifying two kinds of dark matter (DM) component particles at high-energy colliders. The strategy is to notice and distinguish double-peaks(humps) in the missing energy/transverse energy distribution. The relative advantage of looking for {\em missing energy} is pointed out, in view of the fact that the longitudinal component of the momentum imbalance becomes an added input. It thus turns out that an electron-positron collider is better suited for discovering  a two-component DM scenario. Furthermore, using Gaussian fits of 
the distribution histograms, we develop a set of criteria to evaluate the distinguishability of the two-peaks quantitatively.}

\FullConference{The European Physical Society Conference on High Energy Physics (EPS-HEP2023)\\
 21-25 August 2023\\
Hamburg, Germany\\}

\def\mdm{m_{\rm DM}}
\def\mdma{m_{\rm DM_1}}
\def\mdmb{m_{\rm DM_2}}

\newcommand{\bea}{\begin{eqnarray}}
\newcommand{\eea}{\end{eqnarray}}

\renewcommand{\hookAfterAbstract}{%
\par\bigskip

}

\begin{document}
\maketitle
\section{Introduction}

\noindent
Evidence for dark matter (DM) has accumulated from different astrophysical observations like rotation curves of galaxies \cite{Rubin:1970zza,Zwicky:1937zza}, 
gravitational lensing effects around bullet clusters \cite{Hayashi:2006kw}, and cosmological observations like the anisotropy of cosmic microwave background radiation 
(CMBR) \cite{Hu:2001bc} in WMAP \cite{Hinshaw:2012aka,Spergel:2006hy} or PLANCK \cite{Planck:2018vyg} data. Observations further suggest that DM constitutes a 
large portion $(\simeq 24\%)$ of the energy budget of the universe, often expressed in terms of relic density $\Omega h^2\simeq 0.12$ \cite{Planck:2018vyg}, where 
$\Omega={\rho}/{\rho_c}$ represents the cosmological density with $\rho$ being the DM density, $\rho_c$ the critical density and $h$ represents Hubble expansion rate 
in units of 100 km/s/Mpc. However, direct evidence of DM in reproducible terrestrial observations is yet to be found.
We are thus still unable to confirm whether DM consists of elementary particles of the weakly or feebly interacting
types. All one can say with certainty is that neutrinos cannot be the dominant components of DM, and thus
physics beyond the standard model (SM) have to be there if particle DM exists.

Two major classes of ideas, both of which can account for correct relic density, are often discussed. The first category consists of 
weakly interacting massive particle (WIMP) scenarios where the DM particles were in thermal and chemical equilibrium in early universe, 
and have frozen out  when their annihilation rate dropped below the Hubble expansion rate \cite{Bertone:2004pz,Roszkowski:2017nbc,Kolb:1990vq}. 
In the second category, one can have feebly interacting massive particles (FIMP)\cite{Hall:2009bx} which do not thermalise with
the cosmic bath, and are presumably produced from the decay or scattering of some massive particles in thermal bath. 
We shall be concerned here with WIMPs, since they are the likeliest one to be detected in collider experiments which constitute the theme of this paper.

It is of course possible to have more than one DM components simultaneously, and this is the possibility we are concerned with. 
While many of the existing multicomponent DM studies are in the context of WIMPs \cite{Aoki:2012ub,Liu:2011aa,Cao:2007fy,Bhattacharya:2013hva,Esch:2014jpa,Karam:2016rsz,Ahmed:2017dbb,Poulin:2018kap,Aoki:2018gjf,YaserAyazi:2018lrv,Aoki:2017eqn}, scenarios with more than one DM types have   
also been studied \cite{Bhattacharya:2021rwh,DuttaBanik:2016jzv,Choi:2021yps}. 
Direct search experiments \cite{XENON:2018voc,PandaX-4T:2021bab} might probe two component WIMP frameworks via observation 
of a kink in the recoil energy spectrum \cite{Herrero-Garcia:2017vrl,Herrero-Garcia:2018qnz}. We devote the present
discussion to the collider detectability and distinguishability of two DM components, both of whom are of the WIMP type. 
However, studies on collider searches are relatively fewer \cite{Hernandez-Sanchez:2020aop,Konar:2009qr}. 
Here we develop some criteria for the discrimination of two peaks in missing energy distributions, in an illustrative scenario 
where the DM components are pair-produced in cascades along with visible particles, as well DM produced in mono-X topology.
In both cases, we consider cases where  the two DM components belong to two separate
`dark sectors'.
Our purpose is to maximise the visibility
of the two peaks, via discriminants based on their heights, separation and spreads.

We also emphasize that the main principle(s), on which our suggested method of analysis is based,
do not depend on the model used here for illustration. Certain features of the model are course best suited
for substantial production of the two kinds of DM particles, and the decay chains that occur here
simplify and facilitate what we wish to demonstrate. More complicated avenues of DM production are
of course within our horizon, but the points we make here serve as leitmotifs in any analysis. 

We establish further that an electron-positron collider is in most cases better suited for thus 
discerning the two DM components, as compared to hadronic machines. The main reason, as we shall show, 
is that in electron-positron collisions, the full kinematic information, especially that on longitudinal
components of momenta (including missing momenta) can be utilised. In addition, it helps the suppression
of standard model (SM) backgrounds. The option of using polarised beams, too, can be of advantage.

\section{WIMP signal at colliders}
\label{sec:WIMPsignal}
\noindent
WIMP's are produced at colliders either via electroweak hard scattering processes or in weak decays
of other particles.  But no component of
currently designed collider detectors is equipped to register their presence. Their smoking gun signatures,
therefore, result from energy/momentum imbalance in the final state, rising above SM backgrounds
as well as the imbalance due to mis-measurement. Such imbalance can be 
quantified in terms of the following kinematic variables:

\begin{itemize}
\item {\it Missing Transverse Energy or MET ($\slashed{E}_T$)},  defined as:
\bea
\slashed{E}_T = -\sqrt{(\sum_{\ell,j,\gamma} p_x)^2+(\sum_{\ell,j,\gamma} p_y)^2};
\label{eq:MET}
\eea
where the sum runs over all visible objects that include leptons ($\ell$), photons ($\gamma$), jets ($j$), and also unclustered components. 

\item{\it Missing Energy or ME ($\slashed{E}$)}   with respect to the centre-of-mass (CM) energy ($\sqrt{s}$),
defined as:  
\bea
\slashed{E}=\sqrt{s}-\sum_{\ell,j,\gamma} E_{\rm vis}\,;
\label{eq:ME}
\eea
where the sum runs over visible objects like $\ell, j,\gamma$ and unclustered components.

\item {\it Missing Mass or MM ($\slashed{M}$)}, defined as:
\bea
\slashed{M}^2=\left(\sum_ip_{i}-\sum_f p_f\right)^2 \,,
\label{eq:MM}
\eea
\noindent which requires the knowledge of initial state four momenta ($p_i$) and final state ones ($p_f$), where $f$ runs over all the 
visible particles. For mono-photon process $e^+ e^- \to \chi \bar\chi \gamma$, where $\chi(\bar\chi)$ are DM, 
$\slashed{M}^2=s-2\sqrt{s} E_\gamma$. Here, $E_\gamma$ is the energy of the outgoing photon.
\end{itemize}


\noindent
$\slashed{E}$ and $\slashed{M}$ are measurable in $e^+e^-$ (or $\mu^+\mu^-$) machines while
hadron colliders can only measure $\slashed{E_T}$. The scalar sum of transverse momentum, sometimes 
referred as {\it Effective mass} ($H_T$)  is another variable of interest for hadron colliders. 
$\slashed{E}$ or $\slashed{E_T}$ are reconstructible from the energies and momenta of visible particles, 
against which the DM particle(s) recoil.  The resulting signals can be

\medskip

\noindent
$\bullet$ {\bf $n$-leptons + $m$-jets + $p$ photons + $\slashed{E}~(\slashed{E_T}$)} or\\
$\bullet$ {\bf mono-X + $\slashed{E}~(\slashed{E_T}$)}, where X is a jet, a photon, a weak boson or a Higgs. \\


\medskip

\section{Results : di-lepton + $\slashed{E}~(\slashed{E_T})$}

\noindent
We first present the results for di-lepton + $\slashed{E}~(\slashed{E_T})$ final state topology.

\subsection{$\slashed{E_T}$ vs $\slashed{E}$}
\noindent
Here we focus on 
a simple situation having two scalar DM particles with masses $\mdma,\mdmb$ and mass splitting with the corresponding intermediate unstable heavy state (decaying to DM and SM final states) as 
${\Delta m}_1~{\rm and}~ {\Delta m}_2$ respectively. Subsequent $\slashed{E}$ and $\slashed{E_T}$ distributions for different choices of 
$\mdm$ and $\Delta m$ are studied.

\begin{figure}[!hptb]
	\centering
	\subfloat[]{\includegraphics[width=6cm,height=5cm]{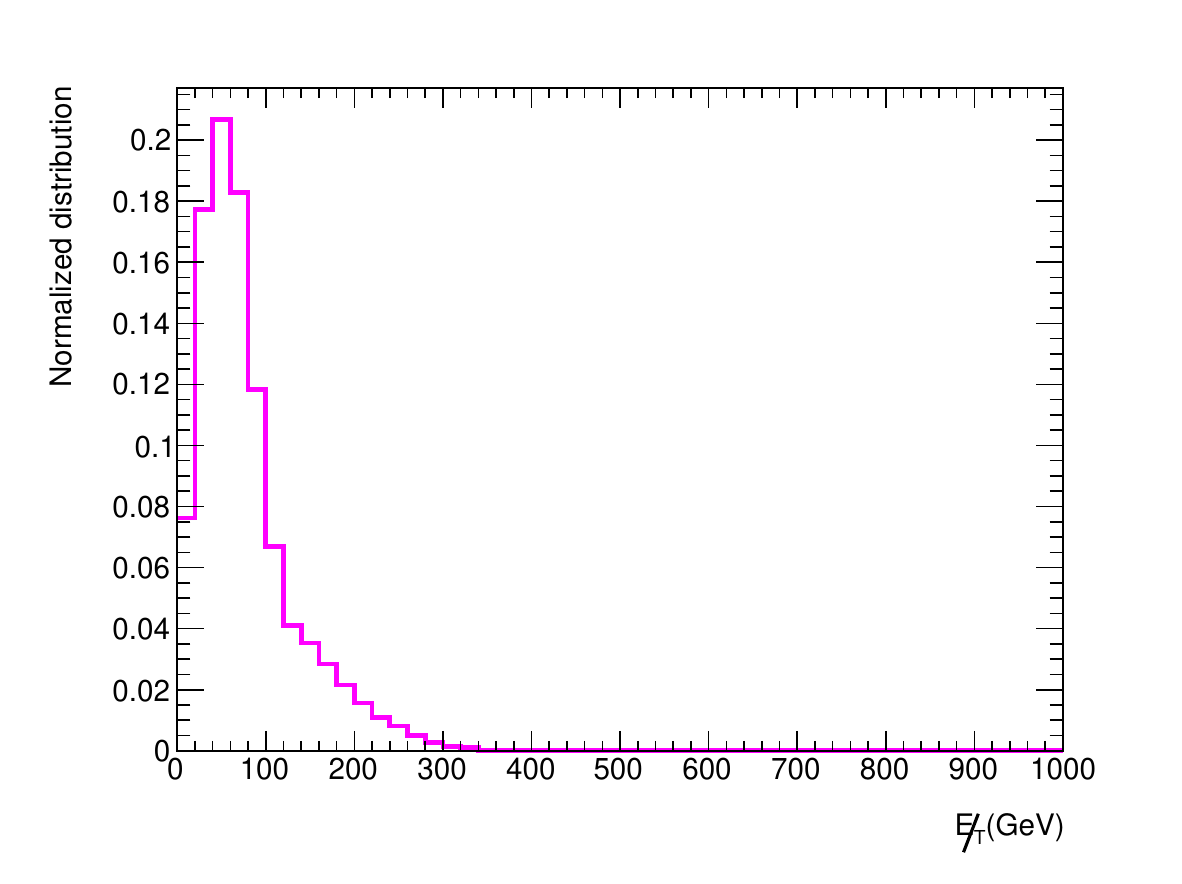}} 
        \subfloat[]{\includegraphics[width=6cm,height=5cm]{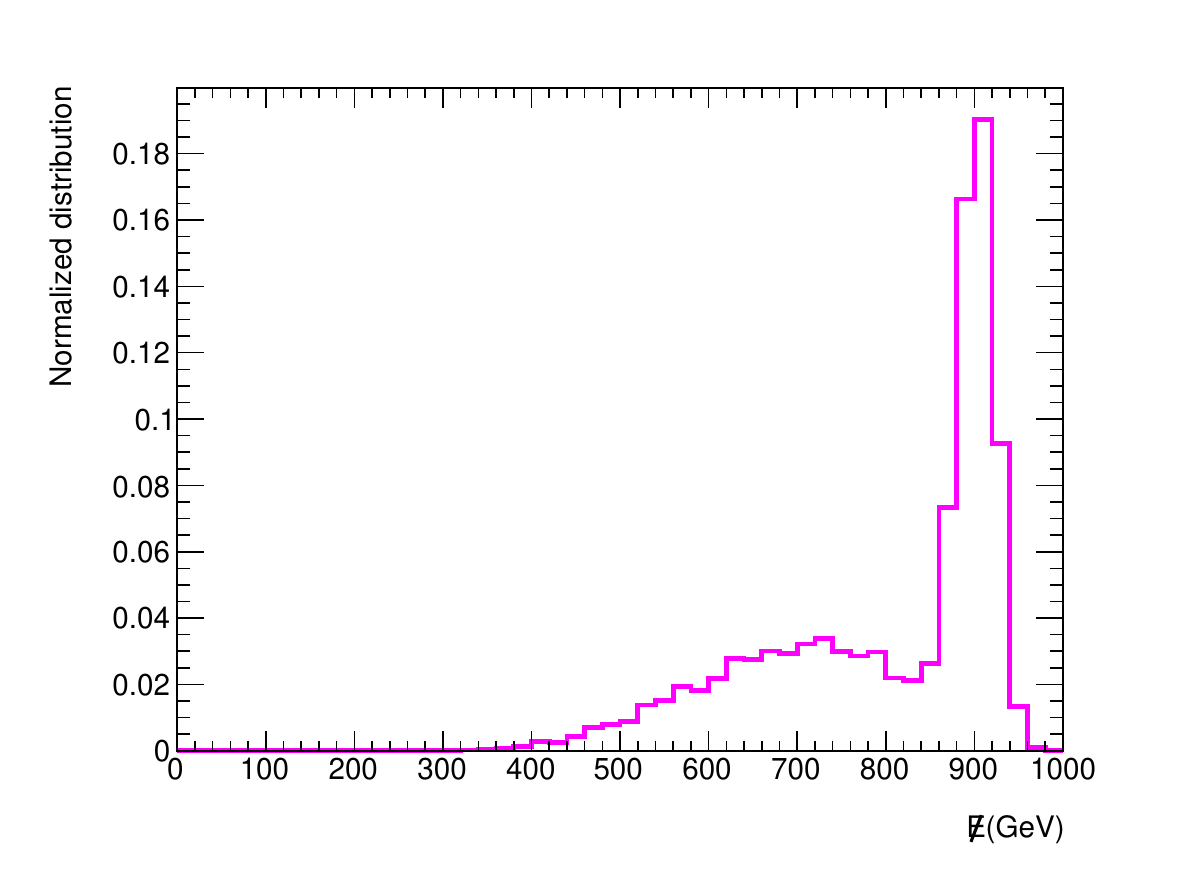}}
	\caption{ Comparison between $\slashed{E_T}$(a) and $\slashed{E}$(b). $\{\mdma, \mdmb, {\Delta m}_1, {\Delta m}_2\}=\{100,100,10,380\}$ GeV. 
	}
	\label{comparison3}
\end{figure}

\noindent
It is clear from Figure~\ref{comparison3}, $\slashed{E}$ distribution is much more sensitive to the two-peak behavior of two-component DM scenario compared to $\slashed{E_T}$. The major reason being the explicit dependence of $\slashed{E}$ on the DM mass, in contrast to $\slashed{E_T}$. This immediately puts a lepton collider like ILC to advantage compared to LHC in this context.

\subsection{Background suppression : Effect of polarization and lepton energy cut}

\noindent
The dominant background contribution to di-lepton + $\slashed{E}~(\slashed{E_T})$ final state comes from $e^+ e^- \rightarrow W^+ W^-$ production, particularly via $t$-channel neutrino exchange. This process involves left-handed electrons and right-handed positrons.
This background can be 
suppressed with maximally right polarized $e^-$ beam and maximally left polarised $e^+$ beam.
Another large background contribution coming from $e^+e^- \rightarrow \nu\bar\nu Z(\ell^+\ell^-)$ can be easily reduced by choosing a suitable polarization and demanding the invariant mass of the lepton pair is outside $m_Z \pm 20$ GeV. 

Furthermore, the lepton energy distribution shows a complementarity with $\slashed{E}$ distribution. 
A cut on the energy of the leading lepton suppresses the background and also helps us 
retain the Gaussian nature of the resulting distribution. However, the choice of the cut is crucial. We present the signal and background distributions corresponding to a benchmark (BP1 in Table~\ref{tab:dm}) for specific polarization choice separately first and then combined after applying the lepton energy cut. 

\begin{table}

{\scriptsize{ \begin{tabular}{|c|c|c|c|c|c|c|c|}
\hline
\makecell{BPs } & \makecell{DM$_1$ \\ $\{m_{\phi^0},~\Delta m_1\}$ } & \makecell{DM$_2$ \\ $\{m_{\psi_1},~\Delta m_2\}$ } & $\Omega_{\phi^0} h^2$    &  $\Omega_{\psi_1} h^2$  & $\sigma_{\phi^0}^{\rm eff}$ (cm$^{2}$) &  $\sigma_{\psi_1}^{\rm eff}$ (cm$^{2}$) & \rm{BR}($H_{\rm inv}$)$\%$\\  \hline \hline 

BP1& $100$ GeV,$~ 10$ GeV & $60.5$ GeV,$~ 370$ GeV & $0.00221$ & $0.1195$ & $3.45 \times 10^{-46}$ & $2.03 \times 10^{-47}$ & $0.25$ \\ \hline
\hline
\end{tabular}}}

\caption{Benchmark point, contribution to relic density, spin-independent direct detection cross-section as well as that of invisible Higgs decay 
branching ratios of the DM components $\phi^0$ and $\psi_1$ are mentioned.}
    \label{tab:dm}
\end{table}

\begin{figure}[!hptb]
	\centering
        	\subfloat[]{\includegraphics[width=6cm,height=5cm]{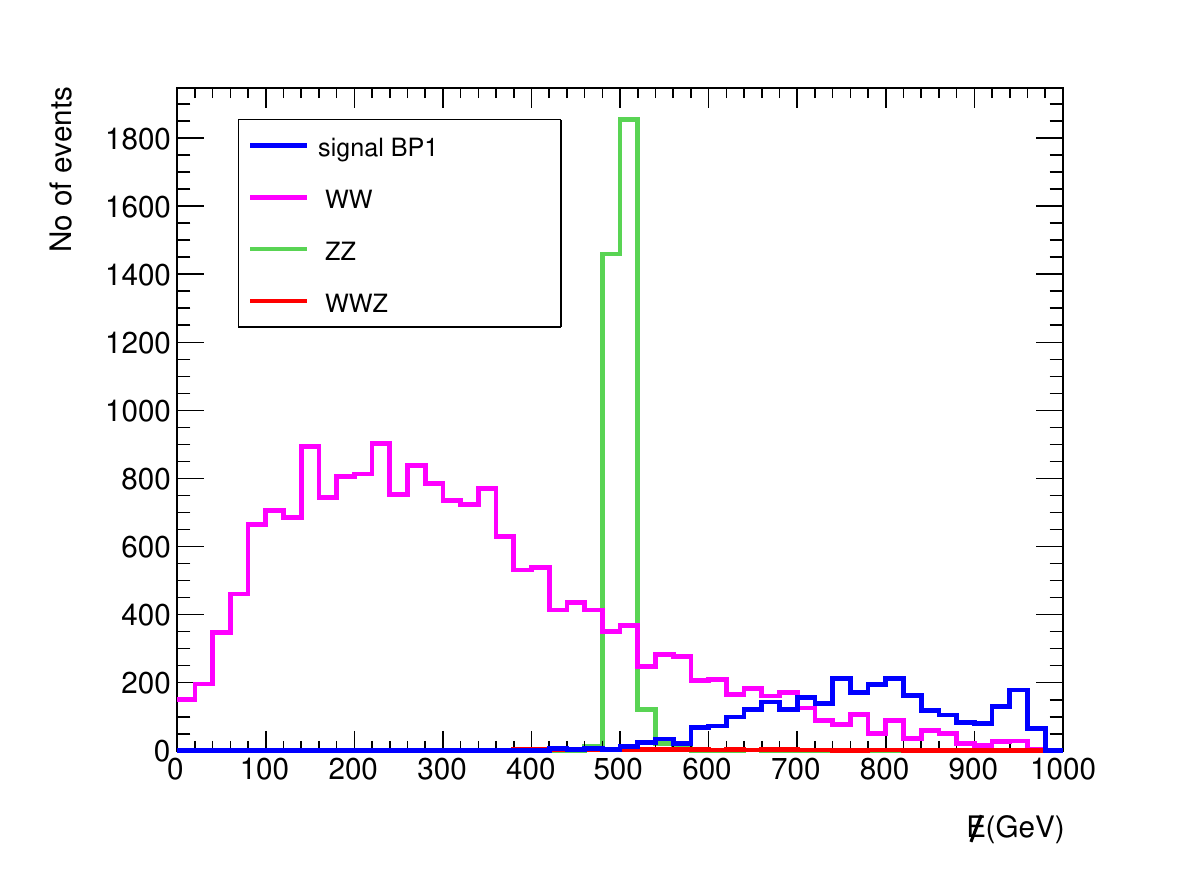}}
	\subfloat[]{\includegraphics[width=6cm,height=5cm]{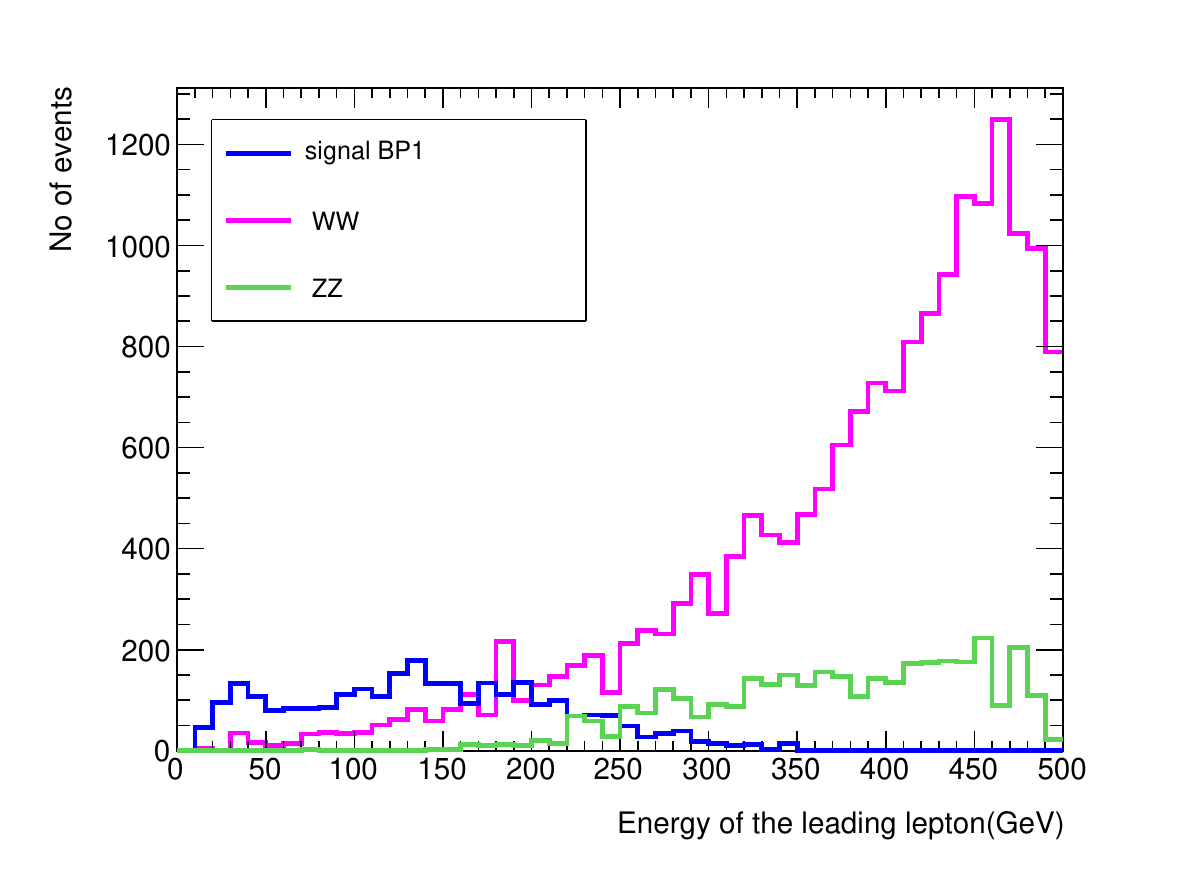}} \\
        \subfloat[]{\includegraphics[width=6cm,height=5cm]{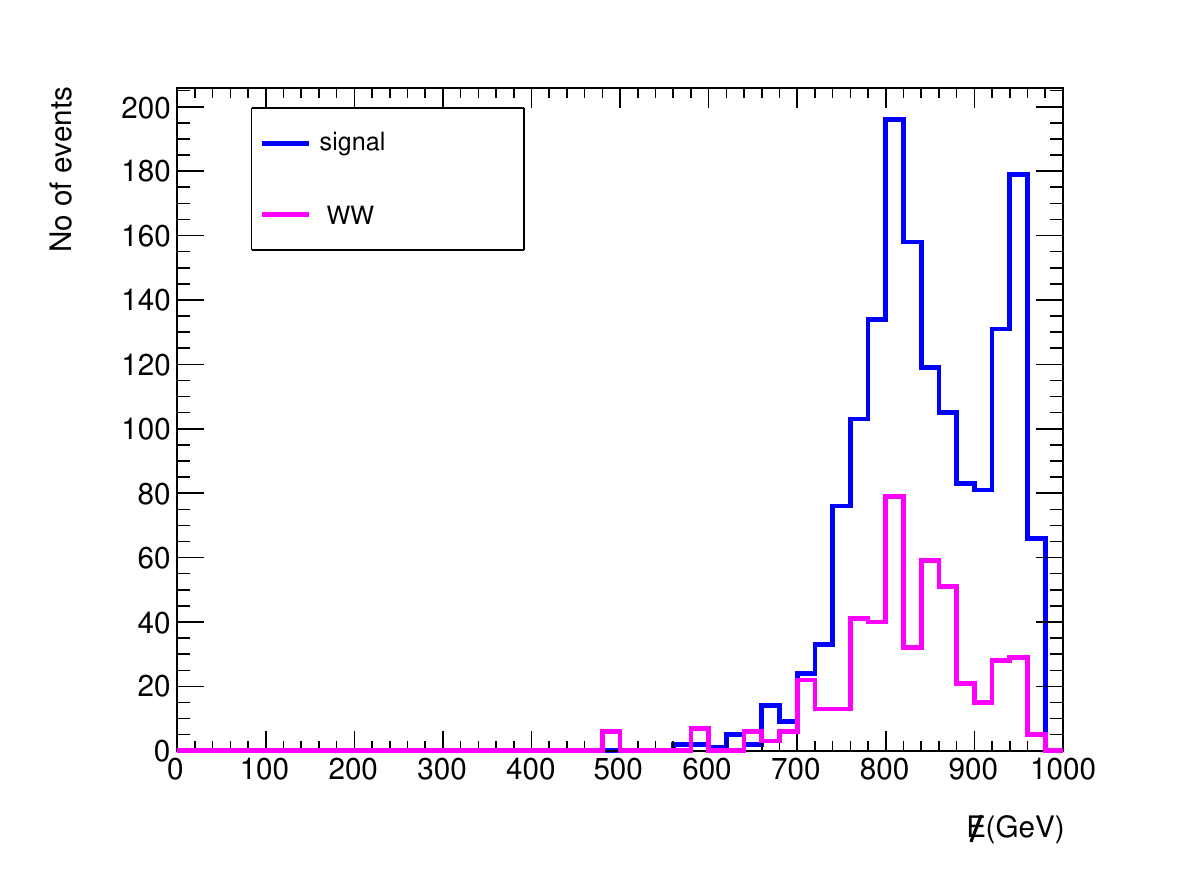}}
        \subfloat[]{\includegraphics[width=6cm,height=5cm]{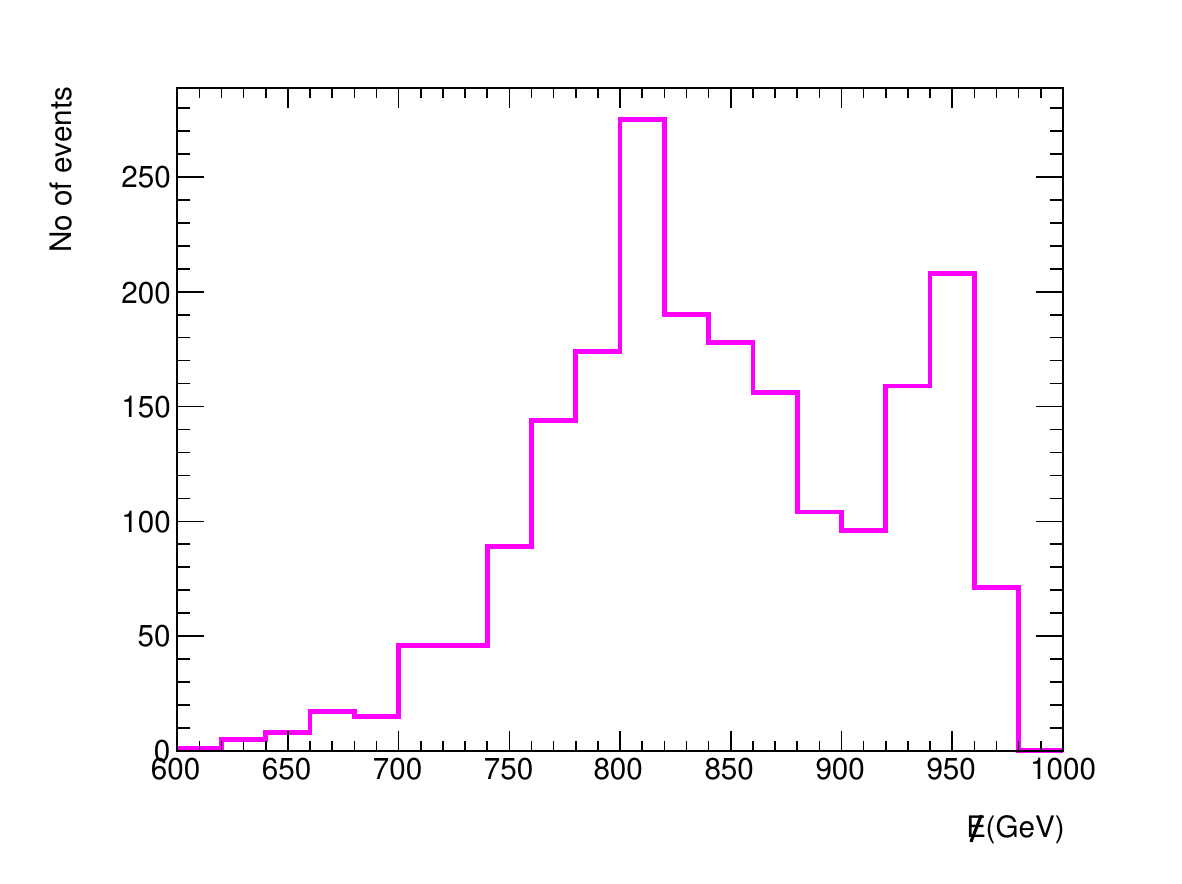}}
	\caption{(a)$\slashed{E}$, (b) $E_{\ell}$ distribution for signal and backgrounds, (c) $\slashed{E}$ distribution for signal and background with $E_{\ell}<$ 150 GeV, (d) $\slashed{E}$ distribution for combined signal + background with $E_{\ell}<$ 150 GeV. $\sqrt{s}=$ 1 TeV, $\{P_{e^{-}}: +0.8, P_{e^{+}}: -0.3\}$ at $\mathcal{L}=$1000 ${\rm fb}^{-1}$. }
	\label{bp1}
\end{figure}

\section{Results : mono-X + $\slashed{E}$}

\noindent
Here too, our study is centred around double bump hunting in $\slashed{E}$ distributions coming from two DM components. In order to explore the features of two-component DM, in the mono-X final state, in a model-independent manner we take up an Effective Field Theory 
(EFT) approach as it suffices to address the DM saturation as well as detectability.

\begin{figure}[!hptb]
	\centering
	\subfloat[]{\includegraphics[width=6cm,height=5cm]{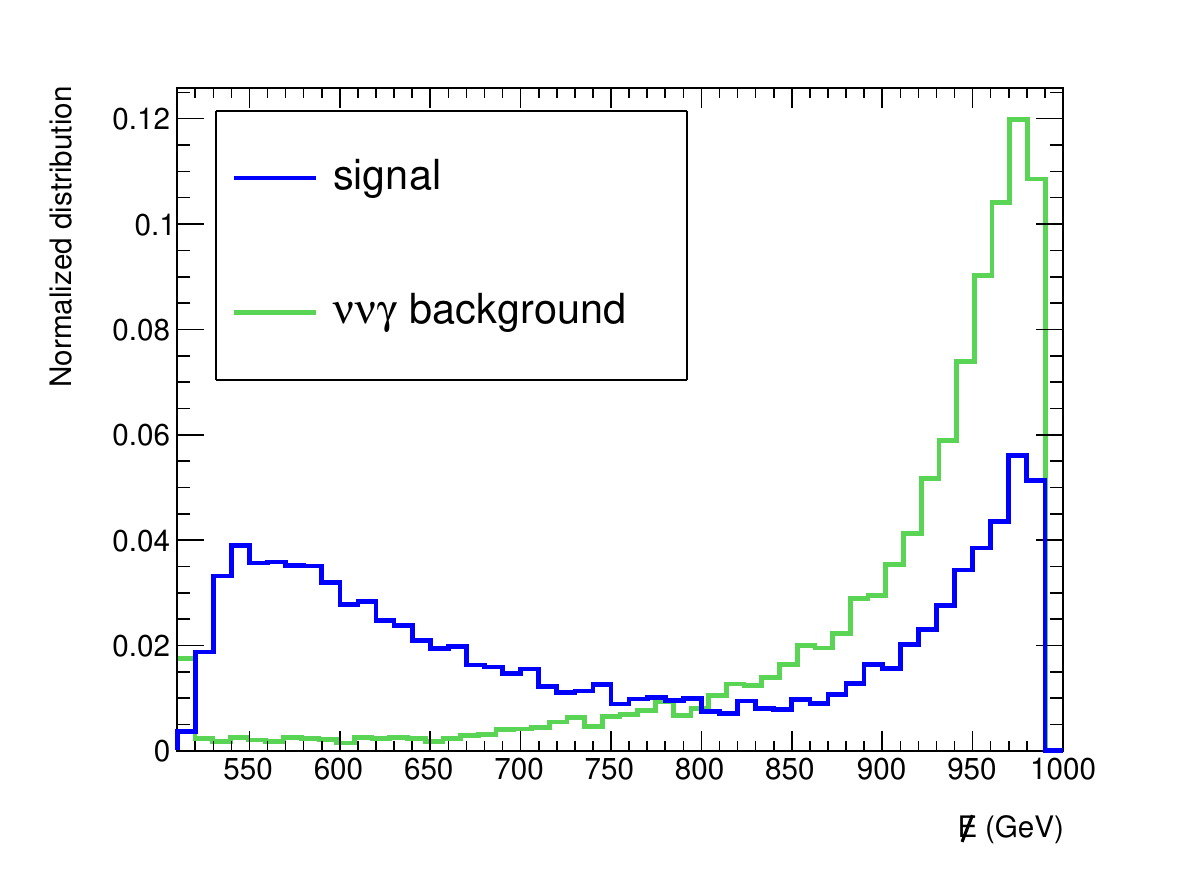}}
        \subfloat[]{\includegraphics[width=6cm,height=5cm]{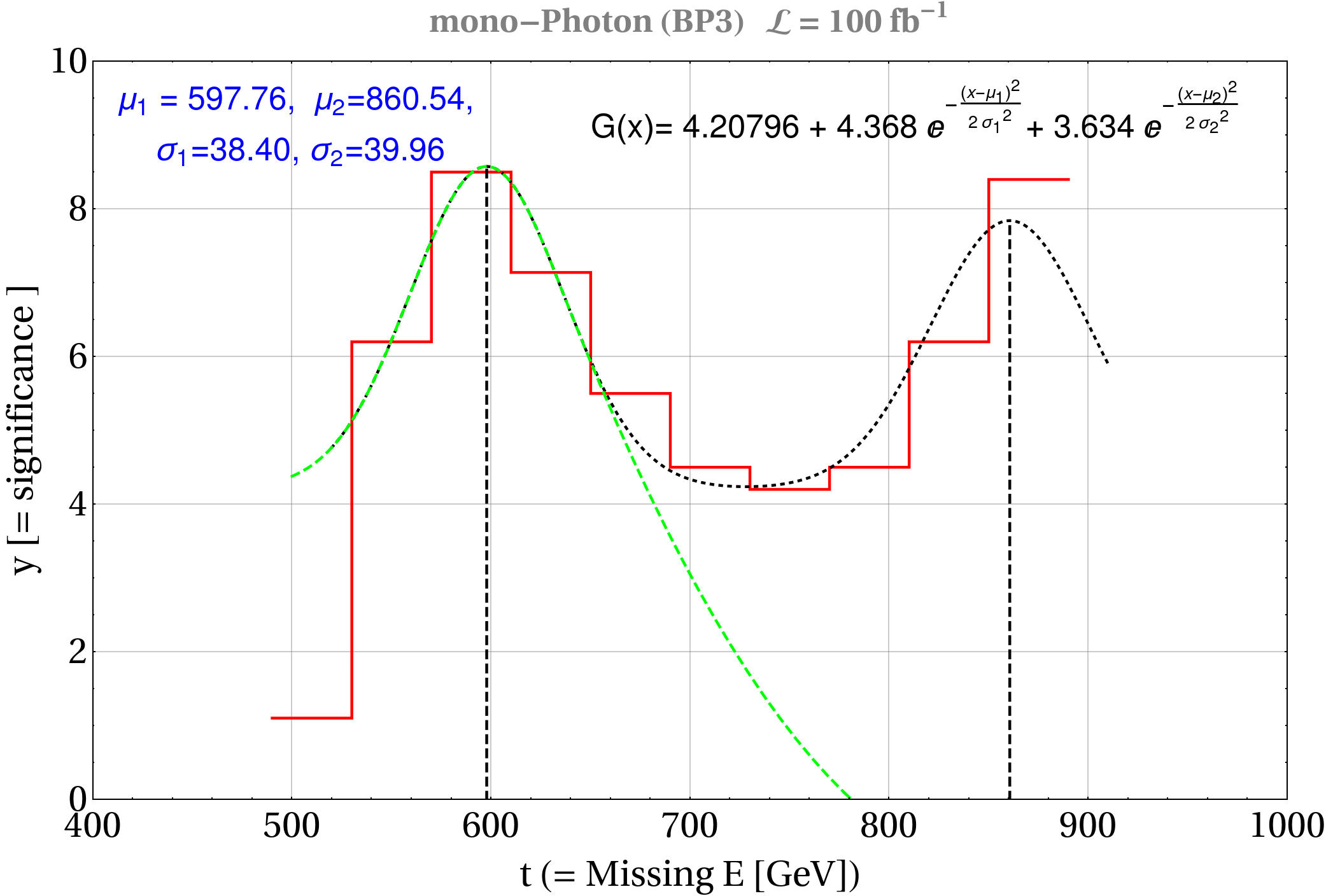}}
	\caption{(a)$\slashed{E}$ distribution for mono-photon final state from $\frac{1}{\Lambda^3} (B_{\mu\nu}B^{\mu\nu} + W_{\mu\nu}W^{\mu\nu})(\bar \chi_1 \chi_1)$ and $\frac{1}{\Lambda^3} (\bar L \Phi \ell_R) (\bar \chi_2 \chi_2)$ with DM masses 100 and 200 GeV respectively, along with $\nu\bar \nu\gamma$ background. (b) Corresponding bin-by-bin local significance distribution.}
	\label{monox_bp1}
\end{figure}

In Figure~\ref{monox_bp1}(a), we show the signal (mono-$\gamma + \slashed{E}$) and background($\nu\bar \nu\gamma$) distribution. One can see a clear two-peak behavior in the signal distribution. However, in the presence of large irreducible background it is difficult to observe the two-peak behavior in the signal+background distribution in this case, unlike the previous di-lepton+$\slashed{E}$ case. The major reason behind this is, limited and correlated observables in the final states,
make background reduction challenging. Therefore, we look at the bin-by-bin local significance distribution while searching for two-peak behavior in this case(see Figure~\ref{monox_bp1}(b)). We present alongside 
the Gaussian fits of the corresponding histograms after minimizing $\chi^2$, using:
 \bea\label{eq:gf}
 G(t)&=& G_1(t)+G_2(t) + \cal{B} \nonumber\\
 &=& A_1~ e^{-\frac{(t-\mu_1)^2}{2\sigma_1^2}}+A_2 ~ e^{-\frac{(t-\mu_2)^2}{2\sigma_2^2}} + \cal{B} ~.
 \eea
The values of $\mu_1$, $\mu_2$, $\sigma_1$ and $\sigma_2$ in each case are indicated in the figure insets.

\section{Quantifying the distinguishability of two peaks}
\noindent
Having discussed the the kinematical variables showing two peaks in $\slashed{E}$ or bin-by-bin local significance distribution, we take up the question : how well can we distinguish the two peaks? In this context we propose two useful statistical variables, namely, $R_{C_3}$, $R_{C_4}$
defined as, 
\bea
 R_{C3} = \frac{\int_{t_1^-}^{t_1^+} y dt-\int_{t_2^-}^{t_2^+} y dt}{\int_{t_1^-}^{t_1^+} y dt+\int_{t_2^-}^{t_2^+} y dt}\,,~
 R_{C4} = \frac{y(t^\prime) -y(t_{\rm min})}{\sqrt{y(t_{\rm min})}} \,;
\label{c3}
 \eea
 where $y=G(t)$, $t_i^\pm=t_i \pm \Delta t$ with $t_i$ denoting the mean value of $i^{\rm th}$ peak of the two peak gaussian function $G(t)$ and 
 $t_{\rm min}$ denotes the local minima between the two peaks. In the definition of $R_{C_4}$, the parameter $t^\prime$ is specified as 
 \bea\nonumber
 t^\prime=\left\{ 
  \begin{array}{ c l }
    t_2 & \quad \textrm{if } ~y(t_2) < y(t_1) , \\
    t_1 & \quad \textrm{if } ~y(t_2) > y(t_1) .
  \end{array}
\right.
\eea
 
\noindent
$R_{C3}$ reflects a comparative estimate of the number of events in the vicinity of the two peaks, and smaller $R_{C3}$ implies 
that the second peak is more significant relative to the first one. $R_{C4}$ quantifies the presence of the peaks with 
respect to the minimum in between, and its bigger values indicate better distinguishability.

\section{Conclusion}

\noindent
We have suggested some methods of distinguishing two DM components at the collider. Two component DM scenarios, pertaining to various models give rise to double peaks in $\slashed{E}$ or $\slashed{E}_T$ distributions, 
whose identification and segregation constitute the quintessence of our investigation.
We look at two important final states pertaining to DM search at collider, namely di-lepton+$\slashed{E}$(DM from cascade of heavier states) and mono-X. 

In the cascade scenario, the key variables that play a role in producing distinguishable peaks are both of the DM masses ($\mdma,\mdmb$) 
and also their mass-splitting with the corresponding heavier intermediate unstable particles ($\Delta m_1, \Delta m_2$). 
We further demonstrate that, while $\slashed{E}_T$ is the canonical
label of invisible particles at hadron colliders, it is in $\slashed{E}$-distributions that the peaks are 
likely to be more prominent. This is because the DM masses do not play a role in $\slashed{E}_T$, 
while they show up in the $\slashed{E}$-distribution, thus making the peaks more distinct, when
the masses of the two DM-components are well-separated. Thus $e^-e^+$ colliders that have both the DM
components within their kinematic reach emerge as their best hunting grounds. In addition, the absence
of QCD backgrounds as well as the possibility of beam polarisation serves to reduce the background
to the DM signals. 

Mono-X events in an $e^+ e^-$ collider bring in a big challenge in distinguishing between two DM components. 
This is in contrast to DM signals produced in cascades, where the backgrounds are easier to reduce. We also demonstrate that bin-by-bin distributions
in signal significance plotted against $\slashed{E}$ obviates the double-peaking behaviour
which the backgrounds otherwise tend to obliterate.

Finally, we offer some prescriptions for distinguishing the two peaks quantitatively. 
For this purpose, we suggest a set of criteria which quantify the height, sharpness and separability
of one peak relative to the other.


\bibliographystyle{JHEP}
\bibliography{ref}
\end{document}